\documentclass[12pt]{article}
\usepackage{graphicx,verbatim,array,multicol,amsfonts,amsmath,subfigure}
\usepackage{fullpage}
\usepackage{chicago}
\usepackage{color}
\usepackage{epsfig,latexsym,amssymb}
\usepackage{lscape}
\usepackage{bm}

\begin{document}

\title{A note on target distribution ambiguity of likelihood-free samplers}
\author{
	S. A.~Sisson\footnote{School of Mathematics and Statistics, University of New South Wales, Sydney, Australia}
	\footnote{Communicating author: Tel: 612 9385 7027; Email: Scott.Sisson@unsw.edu.au}
	\and G. W.~Peters$^*$\footnote{CSIRO  Sydney, Locked Bag 17, North Ryde, NSW, 1670, Australia}
	\and M.~Briers\footnote{QinetiQ Ltd., Malvern, Worcestershire, WR14 3PS, UK}
	\and  Y. Fan$^{*}$
}
\maketitle

\begin{abstract}
\noindent 
Methods for Bayesian simulation in the presence of computationally intractable likelihood functions are of growing interest.
Termed {\it likelihood-free samplers}, standard simulation algorithms such as Markov chain Monte Carlo have been adapted for this setting. 
%
%
In this article, by presenting generalisations of existing algorithms, we demonstrate that likelihood-free samplers can be ambiguous 
over the form of the target distribution. We also consider the theoretical justification of these samplers. Distinguishing between the forms of the target distribution may have implications for the future development of likelihood-free samplers.
%
%
%

\vspace{0.5cm} \noindent \textbf{Keywords:} Approximate Bayesian
computation; Likelihood-free computation; Rejection Sampling; Markov chain Monte Carlo; Sequential Monte Carlo.
\end{abstract}


\section{Introduction}


Bayesian inference proceeds via the posterior distribution $\pi(\theta|y)\propto f(y|\theta)\pi(\theta)$, the updating of prior information $\pi(\theta)$ for a parameter $\theta\in\Theta$ through the likelihood  function $f(y|\theta)$ after observing data $y\in{\mathcal Y}$. 
Numerical algorithms, such as importance sampling, Markov chain Monte Carlo (MCMC) and sequential Monte Carlo (SMC), 
are commonly employed to draw samples from the posterior $\pi(\theta|y)$.

There is growing interest in posterior simulation
in situations  where the likelihood function is computationally intractable i.e.  $f(y|\theta)$ may not be  numerically evaluated pointwise. 
As a result,  sampling algorithms based on repeated likelihood evaluations
require modification for this task. 
Collectively known as {\it likelihood-free samplers} (and also as {\it approximate Bayesian computation})
these methods have been developed across multiple disciplines and literatures. 
They employ generation of auxiliary datasets under the model as a means to circumvent (intractable) likelihood evaluation.

In this article we present general forms of two likelihood-free models, and extend
earlier likelihood-free samplers (based on rejection sampling and MCMC) to these
models. 
In doing so, we demonstrate that likelihood-free samplers are sometimes ambiguous over the exact form of their target distribution: in particular whether samples are obtained from the joint distribution of model parameters and auxiliary datasets, or from the marginal distribution of model parameters only. The interpretation of the auxiliary datasets is quite distinct in each case: under the joint distribution target they play the role of auxiliary parameters, whereas under the marginal distribution target they are simply a means to approximate the likelihood function under Monte Carlo integration.
It may be important for the future development of likelihood-free samplers to make clear the distinction between the two different forms of target distribution, and the interpretation of the auxiliary datasets.

In Section \ref{section:abc} we 
establish the notation and models underlying likelihood-free methods.
In Section \ref{section:samplers} we consider importance sampling, MCMC and SMC algorithms in turn, and discuss sampler validity and algorithm equivalence under both target distributions.
We conclude with a summary and discussion in Section \ref{section:discussion}.

\section{Models for computationally intractable likelihoods}
\label{section:abc}

In essence, likelihood-free methods 
first reduce the observed data, $y$,  to a low-dimensional vector of summary statistics $t_y=T(y)\in{\mathcal T}$, where $\dim(\theta)\leq\dim(t_y)<<\dim(y)$.
Accordingly, the true posterior $\pi(\theta|y)$ is replaced with a new posterior $\pi(\theta|t_y)$. These are equivalent if $t_y$ is sufficient for $\theta$, and $\pi(\theta|t_y)\approx\pi(\theta|y)$ is an approximation if there is some loss of information through $t_y$.
The new target posterior, $\pi(\theta|t_y)$, still assumed to be computationally intractable, is then
embedded 
within an augmented model from which sampling  is viable. Specifically the joint posterior of the
model parameters $\theta$, and auxiliary data $t\in{\mathcal T}$ given observed data $t_y$ 
is 
\begin{equation}
\label{eqn:joint}
	\pi(\theta,t|t_y)\propto K_h(t_y-t)f(t|\theta)\pi(\theta),
\end{equation}
where 
$t\sim f(t|\theta)$ 
may be interpreted as the vector of summary statistics $t=T(x)$ computed from a dataset simulated according to the model $x\sim f(x|\theta)$. Assuming
such simulation is possible, data-generation under the model, $t\sim f(t|\theta)$, forms the basis of computation in the likelihood-free setting -- see Section \ref{section:samplers}.
The target marginal posterior 
$\pi_M(\theta|t_y)$ for the parameters $\theta$, is then obtained as 
\begin{equation}
\label{eqn:marginal}
	\pi_M(\theta|t_y) = c_M\int_{\mathcal{T}}K_h(t_y-t)f(t|\theta)\pi(\theta) dt
\end{equation}
where $(c_M)^{-1}=\int_{\Theta}\int_{\mathcal{T}}K_h(t_y-t)f(t|\theta)\pi(\theta) dtd\theta$ normalises (\ref{eqn:marginal}) such that it is a density in $\theta$ (e.g. \citeNP{reeves+p05}; \citeNP{wilkinson08}; \citeNP{blum09}; \citeNP{sisson+f10}; \citeNP{fernhead+p10}).
The function $K_h(t_y-t)$ is a standard kernel function, with scale parameter $h\geq 0$, which weights the intractable posterior with high density in regions $t\approx t_y$ where auxiliary and observed datasets are similar.
As such, $\pi_M(\theta|t_y)\approx\pi(\theta|t_y)$ forms an approximation to the intractable posterior via (\ref{eqn:marginal}) through standard smoothing arguments (e.g. \citeNP{blum09}).
In the  case as $h\rightarrow 0$, so that
$K_h(t_y-t)$ becomes a point mass at the origin (i.e. $t_y=t$) and is zero elsewhere,  if $t_y$ is sufficient for $\theta$ then
the intractable posterior marginal $\pi_M(\theta|t_y)=\pi(\theta|t_y)=\pi(\theta|y)$ is recovered exactly (although small $h$ is usually impractical -- see Section \ref{section:samplers}). 
Various choices of smoothing kernel $K$ have been examined  in the literature (e.g. \shortciteNP{marjoram+mpt03}; \shortciteNP{beaumont+zb02}; \shortciteNP{peters+fs08a}; \shortciteNP{peters+nsfy10}; \citeNP{blum09}; \citeNP{sisson+f10}).

For our discussion on likelihood-free samplers, it is convenient to consider
a generalisation of the joint distribution  (\ref{eqn:joint}) incorporating $S\geq 1$ 
auxiliary summary vectors
\begin{equation*}
	\pi_J(\theta,t^{1:S}|t_y)
	 \propto
	\tilde{K}_h(t_y,t^{1:S})f(t^{1:S}|\theta)\pi(\theta)
\end{equation*}
where $t^{1:S}=(t^1,\ldots,t^S)$  and $t^1,\ldots,t^S\sim f(t|\theta)$ are $S$ independent datasets generated from the (intractable) model. As the auxiliary datasets are, by construction, conditionally independent given $\theta$, we have $f(t^{1:S}|\theta)=\prod_{s=1}^Sf(t^s|\theta)$.
We follow \shortciteN{delmoral+dj08} and specify the kernel $\tilde{K}$  as $\tilde{K}_h(t_y,t^{1:S})=S^{-1}\sum_{s=1}^SK_h(t_y-t^s)$,
which produces the joint posterior
\begin{equation}
\label{eqn:jointS}
	\pi_J(\theta,t^{1:S}|t_y)
	 =
	 c_J
	\left[\frac{1}{S}\sum_{s=1}^SK_h(t_y-t^s)\right]
	\left[\prod_{s=1}^Sf(t^s|\theta)\right]
	\pi(\theta),
\end{equation}
with $c_J>0$ the appropriate normalisation constant, 
where in (\ref{eqn:jointS}) we extend the uniform kernel choice of $K(t_y-t^s)$ 
by \shortciteN{delmoral+dj08} 
to the general case. It is easy to see that, by construction,  $\int_{{\mathcal T}^S}\pi_J(\theta,t^{1:S}|t_y)dt^{1:S}=\pi_M(\theta|t_y)$ admits the distribution (\ref{eqn:marginal}) as a marginal distribution (c.f. \shortciteNP{delmoral+dj08}).
The case  $S=1$ with $\pi_J(\theta,t^{1:S}|t_y)=\pi(\theta,t|t_y)$ corresponds to the more usual joint posterior (\ref{eqn:joint}) in the likelihood-free setting.


There are two obvious approaches to posterior simulation from $\pi_M(\theta|t_y)\approx\pi(\theta|t_y)$ as an approximation to $\pi(\theta|y)$. The first approach proceeds by sampling directly on the augmented model $\pi_J(\theta,t^{1:S}|t_y)$, realising joint samples $(\theta,t^{1:S})\in\Theta\times{\mathcal T}^S$ before {\it a posteriori} marginalisation over $t^{1:S}$ (i.e. by discarding the $t^s$ realisations from the sampler output). In this approach, the summary quantities $t^{1:S}$ are treated as parameters in the augmented model.
The second approach is to sample from $\pi_M(\theta|t_y)$ directly, a lower dimensional space, by approximating the integral (\ref{eqn:marginal}) via Monte Carlo integration in lieu of each posterior evaluation of $\pi_M(\theta|t_y)$. 
In this case
\begin{equation}
\label{eqn:abc-monte-carlo}
	\pi_M(\theta|t_y) 
	\propto 
	\pi(\theta)\int_{\mathcal T}K_h(t_y-t)f(t|\theta)dt
	 \approx 
	\frac{\pi(\theta)}{S}\sum_{s=1}^S K_h(t_y-t^{s})
	:=\hat{\pi}_M(\theta | t_y),
\end{equation}
where $t^1,\ldots,t^S\sim f(t|\theta)$.
This expression, 
examined by various 
authors (e.g. \shortciteNP{marjoram+mpt03}; \citeNP{reeves+p05}; \shortciteNP{sisson+ft07}; \shortciteNP{ratman+ahwr08}; \shortciteNP{toni+wsis08}; \shortciteNP{peters+fs08a}),
requires multiple generated datasets $t^1,\ldots,t^ S$, for each evaluation of the
marginal posterior distribution $\pi_{M}(\theta|t_y)$. As with standard Monte Carlo approximations, $\mbox{Var}[\hat{\pi}_M(\theta|t_y)]$ reduces as $S$ increases, with $\lim_{S\rightarrow\infty}\mbox{Var}[\hat{\pi}_M(\theta|t_y)]=0$. For the marginal posterior distribution, the quantities $t^{1:S}$ serve only as a means to estimate $\pi_M(\theta|t_y) $, and do not otherwise enter the model explicitly. The number of samples $S$ directly impacts on the variance of the estimation.

We now examine the relationships between, and technical validity of, likelihood-free samplers constructed with 
$\pi_M(\theta|t_y)$ and $\pi_J(\theta,t^{1:S}|t_y)$ as the target distribution.

\section{Sampler ambiguity and validity}
\label{section:samplers}

In this section we examine each of the basic sampler types: rejection sampling, MCMC and population-based methods.
 We extend the first
two of these algorithms to multiple data generations ($S\geq 1$). 
We will examine sampler validity 
with respect to the two target posterior distributions $\pi_M(\theta|t_y)$ and $\pi_J(\theta,t^{1:S}|t_y)$,
and demonstrate algorithm equivalence under both the joint and marginal distributional
targets.

\subsection{Rejection samplers}
\label{sec:IS}

Rejection-based likelihood-free samplers were
developed in the population genetics literature (\shortciteNP{tavare+bgd97}; \shortciteNP{pritchard+spf99}; \shortciteNP{marjoram+mpt03}).
 Table \ref{table:abc-rej} presents a generalisation of the rejection sampling algorithm. The specific case of $S=1$ is the original implementation of the sampler. We now demonstrate that this algorithm has both 
 $\pi_M(\theta|t_y)$ and 
 $\pi_J(\theta,t^{1:S}|t_y)$ as target distributions.

\begin{table}[htb]
\centering
\begin{tabular}{l}
{\bf LF-REJ Algorithm}\\
\hline
1. Generate $\theta\sim\pi(\theta)$ from the prior.\\
2. Generate $t^1,\ldots,t^S\sim f(t|\theta)$ independently from the model.\\
3. Accept $\theta$ with probability proportional to $\frac{1}{S}\sum_{s=1}^SK_h(t_y-t^s)$.
\vspace{1mm}\\
\hline
\end{tabular}
\caption{\small\label{table:abc-rej} The generalised likelihood-free rejection sampling (LF-REJ) algorithm.
}
\end{table}

We first assume the joint model target $\pi_{J}(\theta,t^{1:S}|t_y)$, under the LF-REJ algorithm. Following Table \ref{table:abc-rej}, a sample $(\theta,t^{1:S})$ is first drawn from the prior predictive distribution $\pi(\theta,t^{1:S})=\pi(\theta)\prod_{s=1}^Sf(t^s|\theta)$ (steps 1 and 2). 
 The acceptance probability (step 3) for 
 $(\theta,t^{1:S})$ under a rejection sampler targeting (\ref{eqn:jointS}) is proportional to
\[
	\frac{\pi_J(\theta,t^{1:S}|t_y)}{\pi(\theta,t^{1:S})}=\frac{1}{S}\sum_{s=1}^SK_h(t_y-t^s)
\]
as indicated in Table \ref{table:abc-rej}.
{\it A posteriori} marginalisation over $t^{1:S}\in\mathcal{T}^S$ (by discarding the $t^{1:S}$ realisations) then provides draws from $\pi_M(\theta|t_y)$.

If we now assume the marginal model target, $\pi_M(\theta|t_y)$, a sample $\theta$ is first drawn from the prior (Table  \ref{table:abc-rej}, step 1). The acceptance probability for this sample is then proportional to $\pi_M(\theta|t_y)/\pi(\theta)$, which via (\ref{eqn:abc-monte-carlo}) is itself approximately proportional to
\[
	\frac{\hat{\pi}_M(\theta|t_y)}{\pi(\theta)}
	= \frac{1}{S}\sum_{s=1}^SK_h(t_y-t^s),
\]
using the Monte Carlo draws $t^1,\ldots,t^S$ from the model (steps 2 and 3).
 Note that while $\hat{\pi}_M(\theta|t_y)/\pi(\theta)$ is an approximation of the acceptance rate, 
 it is unbiased for all $S\geq 1$.
 Thus, while smaller $S$ 
 will result in more variable acceptance probabilities, the accepted samples will still correspond to  draws from $\pi_M(\theta|t_y)$ for all $S\geq 1$.
 Hence, from the above we have that the LF-REJ algorithm successfully targets both $\pi_J(\theta,t^{1:S}|t_y)$ 
 and $\pi_M(\theta|t_y)$, for any $S \geq 1$.

\subsection{Markov chain Monte Carlo samplers}

MCMC-based likelihood-free samplers were introduced to avoid rejection sampling inefficiencies when the posterior and prior were sufficiently different (\shortciteNP{marjoram+mpt03}; \shortciteNP{bortot+cs07}; \citeNP{sisson+f10}). The generalised likelihood-free MCMC algorithm for $S \geq 1$  is presented in Table \ref{table:abc-mcmc}. Again, $S=1$ with a uniform kernel, $K$, is the original implementation of this sampler.

\begin{table}[htb]
\centering
\begin{tabular}{l}
{\bf LF-MCMC Algorithm}\\
\hline
\phantom{1.} Initialise $\theta_1$ (and $t_1^{1:S}=(t_1^{1},\ldots,t_1^{S})$ with $t_1^{s}\sim f(t|\theta_1)$ drawn from the model)\\
At stage $n\geq1$\\
1. Generate $\theta\sim q(\theta_n, \theta)$ from a proposal distribution.\\
2. Generate $t^{1:S}=(t^1,\ldots,t^S)$ with $t^s\sim f(t|\theta)$ drawn independently from the model.
\vspace{1mm}\\
3.  With probability
	$
	\min\left\{1, \frac{\frac{1}{S}\sum_sK_h(t_y-t^s)\pi(\theta)q(\theta,\theta_n)}{\frac{1}{S}\sum_sK_h(t_y-t_n^{s})\pi(\theta_n)q(\theta_n,\theta)}\right\}
	$
	accept $\theta_{n+1}=\theta$, ($t_{n+1}^{1:S}=t^{1:S}$)
	\vspace{1mm}\\
\:\:\:\:\:\:otherwise set $\theta_{n+1}=\theta_n$, ($t_{n+1}^{1:S}=t_n^{1:S}$).\\
4. Increment $n=n+1$ and go to 1.\\
\hline
\end{tabular}
\caption{\small\label{table:abc-mcmc} The generalised likelihood-free MCMC (LF-MCMC) algorithm.
Statements in parentheses involving $t^{1:S}$ relate to sampler with target $\pi_J(\theta,t^{1:S}|t_y)$.
}
\end{table}

The LF-MCMC sampler was introduced in the context of targeting the marginal posterior distribution $\pi_M(\theta|t_y)$.
\shortciteN{marjoram+mpt03} and \shortciteN{wegmann+le09} (for $S=1$)
present variations on proofs of detailed balance under this assumption. We now demonstrate that for finite (i.e. practical values of) $S$, the LF-MCMC sampler is theoretically only valid under the joint posterior target $\pi_J(\theta,t^{1:S}|t_y)$.

Implementing the LF-MCMC sampler assuming the marginal posterior target $\pi_M(\theta|t_y)$, and a proposal density  $q(\theta_n,\theta)$ for $\theta$, the probability of accepting the move from $\theta_n$ at time $n$ to a proposed value $\theta\sim q(\theta_n,\theta)$ is
given by
\begin{equation}
\label{eqn:mcmc1}
	\min\left\{1,\frac{\pi_M(\theta|t_y)q(\theta,\theta_n)}{\pi_M(\theta_n|t_y)q(\theta_n,\theta)}\right\}\approx
	\min\left\{1, \frac{\frac{1}{S}\sum_sK_h(t_y-t^s)\pi(\theta)q(\theta,\theta_n)}{\frac{1}{S}\sum_sK_h(t_y-t_n^{s})\pi(\theta_n)q(\theta_n,\theta)}\right\}
\end{equation}
via  (\ref{eqn:abc-monte-carlo}).
Unlike rejection sampling, where the acceptance probability is proportional to an unbiased estimate  $\hat{\pi}_M(\theta |t_y)/\pi(\theta)$,  the above Markov chain acceptance probability consists of a ratio of two unbiased estimates $\hat{\pi}_M(\theta  | t_y)/\hat{\pi}_M(\theta_n | t_y)$. As such, the estimate of the acceptance probability (involving this ratio) is biased, as in general $\mathbb{E}[X/Y]\neq \mathbb{E}[X]/\mathbb{E}[Y]$.
Only as $S\rightarrow\infty$ so that the bias of the ratio diminishes, can this algorithm target the marginal posterior $\pi_M(\theta|t_y)$.
Many authors (e.g. \shortciteNP{marjoram+mpt03}; \shortciteNP{bortot+cs07}; \shortciteNP{wegmann+le09} and others) implement the LF-MCMC algorithm with $S=1$, which by this argument appears too small to result in 
an unbiased sampler targeting $\pi_M(\theta|t_y)$.

If we now consider an MCMC algorithm targeting the joint posterior $\pi_{J}(\theta,t^{1:S}|t_y)$, taking the specific form in Equation (\ref{eqn:jointS}), the probability of accepting a proposed move from $(\theta_n,t_n^{1:S})$ at time $n$ to 
\[
	(\theta,t^{1:S})\sim q[(\theta_n,t_n^{1:S}),(\theta,t^{1:S})]=q(\theta_n,\theta)\prod_{s=1}^Sf(t^s|\theta)
\]
at time $n+1$, is then
\begin{equation}
\label{eqn:mcmc2}
	\min\left\{1, \frac{\pi_J(\theta,t^{1:S}|t_y)q[(\theta,t^{1:S}),(\theta_n,t_n^{1:S})]}{\pi_J(\theta_n,t_n^{1:S}|t_y)q[(\theta_n,t_n^{1:S}),(\theta,t^{1:S})]}\right\} 
	=
	\min\left\{1, \frac{\frac{1}{S}\sum_sK_h(t_y-t^s)\pi(\theta)q(\theta,\theta_n)}{\frac{1}{S}\sum_sK_h(t_y-t_n^{s})\pi(\theta_n)q(\theta_n,\theta)}\right\}.
\end{equation}
This sampler correctly targets $\pi_J(\theta,t^{1:S}|t_y)$ by construction, and the acceptance probability  (\ref{eqn:mcmc2}) is exact.

Hence, through the equivalence of the acceptance probabilities (\ref{eqn:mcmc1}) and (\ref{eqn:mcmc2}), the auxiliary variable LF-MCMC sampler targeting $\pi_J(\theta,t^{1:S}|t_y)$ results in exactly the same algorithm as an LF-MCMC sampler targeting $\pi_M(\theta|t_y)$, for {\it any} $S\geq 1$.
Thus, despite the above argument of bias in marginal samplers for finite $S$, implementations of marginal LF-MCMC samplers are in practice unbiased for $S\geq 1$, in that the 
sampler must correctly
produce draws from $\pi_M(\theta|t_y)$. 
However, this practical unbiasedness is
strictly only available through that conveyed by the equivalent sampler targeting 
$\pi_J(\theta,t^{1:S}|t_y)$.

\subsection{Population-based samplers }

Population-based likelihood-free samplers  were introduced to circumvent poor mixing in  MCMC samplers (\shortciteNP{sisson+ft07};  \shortciteNP{toni+wsis08}; \shortciteNP{beaumont+cmr08}; \shortciteNP{peters+fs08a}; \shortciteNP{delmoral+dj08}). 
These samplers propagate a population of {\it particles}, $\theta^{(1)},\ldots,\theta^{(N)}$, with associated importance weights $W(\theta^{(i)})$, through a sequence of related densities $\phi_1(\theta_1),\ldots,\phi_{n}(\theta_n)$,
which defines a smooth transition from the distribution $\phi_1$, from which direct sampling is available, 
to $\phi_n$ the target distribution. 
For likelihood-free samplers, $\phi_k$
is defined by allowing $K_{h_k}(t_y-t)$ to place greater density on regions for which $t_y\approx t$ as $k$ increases (that is, the bandwidth $h_k$ decreases with $k$). 
Hence,  we denote $\pi_{J,k}(\theta, t^{1:S}|t_y)\propto \tilde{K}_{h_k}(t_y,t^{1:S})f(t^{1:S}|\theta)\pi(\theta)$ and $\pi_{M,k}(\theta|t_y)\propto\pi(\theta)\int_{\mathcal{T}^S}\tilde{K}_{h_k}(t_y,t^{1:S})f(t^{1:S}|\theta)dt^{1:S}$ 
for $k=1,\ldots,n$,
 under the joint and marginal posterior models respectively.

\subsubsection{Sequential Monte Carlo-based samplers}

Under the sequential Monte Carlo samplers algorithm \shortcite{delmoral+dj06}
the particle population $\theta_{k-1}$ drawn from the distribution $\phi_{k-1}(\theta_{k-1})$ at time $k-1$ is mutated to $\phi_k(\theta_k)$ by the kernel $M_k(\theta_{k-1},\theta_k)$. 
The weights 
for the mutated particles $\theta_k$
may be obtained
as
$
	W_k(\theta_k) = W_{k-1}(\theta_{k-1})w_{k}\left(\theta_{k-1}, \theta_k\right)
$
where, for the marginal model sequence $\pi_{M,k}(\theta_k|t_y)$, the incremental weight is 
\begin{equation}
\label{eqn:smc-weight-increment}
	w_{k}\left( \theta_{k-1}, \theta_k\right) 
	= 
	\frac{\pi_{M, k}(\theta_{k}|t_y) L_{k-1}\left( \theta_{k},\theta_{k-1}\right) }{\pi _{M, k-1}(\theta_{k-1}|t_y) M_{k}\left( \theta_{k-1},\theta_{k}\right) }
	\approx
	\frac{\hat{\pi}_{M, k}(\theta_{k}|t_y) L_{k-1}\left( \theta_{k},\theta_{k-1}\right) }{\hat{\pi} _{M, k-1}(\theta_{k-1}|t_y) M_{k}\left( \theta_{k-1},\theta_{k}\right) },
\end{equation}
where, following (\ref{eqn:abc-monte-carlo}), 
\begin{equation*}
	\hat{\pi}_{M, k}(\theta_{k}|t_y):=\frac{\pi(\theta)}{S}\sum_{s=1}^S K_{h_k}(t_y-t^{s})
\end{equation*}
is proportional to an (unbiased) estimate of $\pi_{M, k}(\theta_{k}|t_y)$ based on $S$ Monte Carlo draws $t^1,\ldots,t^S\sim f(t|\theta_k)$.
Here $L_{k-1}\left( \theta_{k},\theta_{k-1}\right)$ is 
a reverse-time kernel describing the mutation of particles from $\phi_k(\theta_k)$ at time $k$ to $\phi_{k-1}(\theta_{k-1})$ at time $k-1$. As with the LF-MCMC algorithm, 
the incremental weight (\ref{eqn:smc-weight-increment}) consists of the ``biased'' ratio $\hat{\pi}_{M,k}(\theta_k|t_y)/\hat{\pi}_{k-1}(\theta_{M,k-1}|t_y)$ for finite $S\geq1$.
  
If we now consider a sequential Monte Carlo sampler under the joint model $\pi_{J, k}(\theta, t^{1:S}|t_y)$, with the natural mutation kernel factorisation
\[
	M_k[(\theta_{k-1},t_{k-1}^{1:S}),(\theta_{k},t_k^{1:S})]=M_k(\theta_{k-1},\theta_k)\prod_{s=1}^Sf(t_k^{s}|t_y)
\]
(and similarly for $L_{k-1}$),  following the form of (\ref{eqn:smc-weight-increment}), the incremental weight is exactly
\begin{equation}
\label{eqn:smc-weight-increment-J}
	w_{k}\left[ (\theta_{k-1}, t_{k-1}^{1:S}), (\theta_k, t_k^{1:S})\right] 
	= 
	\frac{\frac{1}{S}\sum_sK_{h_k}(t_y-t_k^{s})\pi(\theta_k) L_{k-1}\left( \theta_{k},\theta_{k-1}\right)}{\frac{1}{S}\sum_sK_{h_{k-1}}(t_y-t_{k-1}^{s})\pi(\theta_{k-1}) M_{k}\left( \theta_{k-1},\theta_{k}\right)}.
\end{equation}
Hence, as the incremental weights (\ref{eqn:smc-weight-increment}, \ref{eqn:smc-weight-increment-J}) are equivalent, they induce identical SMC algorithms  for both marginal and joint models $\pi_{M}(\theta|t_y)$ and $\pi_{J}(\theta,t^{1:S}|t_y)$. 
%
As a result, while applications of the marginal sampler targeting $\pi_{M}(\theta|y)$ are theoretically biased for finite $S\geq 1$, as before, they are in practice unbiased through association with the equivalent sampler on joint space targeting $\pi_{J}(\theta,t^{1:S}|t_y)$.

We note that a theoretically unbiased sampler targeting $\pi_M(\theta|t_y)$, for all $S\geq 1$, can be obtained by careful choice of the kernel $L_{k-1}(\theta_k,\theta_{k-1})$. 
For example,
\shortciteN{peters+fs08a} use the suboptimal kernel  (\shortciteNP{delmoral+dj06})
\begin{equation}
\label{eqn:L3}
    L_{k-1}(\theta_k, \theta_{k-1})=\frac{\pi_{M,k-1}(\theta_{k-1}|t_y)M_k(\theta_{k-1},
    \theta_k)}
        {\int\pi_{M, k-1}(\theta_{k-1}|t_y)M_k(\theta_{k-1},\theta_k)d\theta_{k-1}},
\end{equation}
from which
 the incremental weight 
(\ref{eqn:smc-weight-increment}) is approximated by
\begin{eqnarray}
\label{eqn:L3app}
    w_k(\theta_{k-1},\theta_k)
    & = &
    \pi_{M,k}(\theta_k|t_y)/\int \pi_{M, k-1}(\theta_{k-1}|t_y)M_k(\theta_{k-1},\theta_k)d\theta_{k-1}\nonumber\\
    & \approx & 
    \hat{\pi}_{M, k}(\theta_k|t_y)/\sum_{i=1}^N W_{k-1}(\theta^{(i)}_{k-1})M_k(\theta^{(i)}_{k-1},\theta_k) .
\end{eqnarray}
Under this choice of backward kernel, the weight calculation 
is now unbiased for all $S \geq 1$, 
since the approximation $\hat{\pi}_{M,k-1}(\theta|y)$ in the denominator of (\ref{eqn:smc-weight-increment}) is no longer needed.

In practice, application of SMC samplers in the likelihood-free setting requires the avoidance of severe particle depletion.
Targeting $\pi_{J}(\theta,t^{1:S}|t_y)$, \shortciteN{delmoral+dj08} use a standard MCMC kernel in combination with a large number of slowly changing distributions $\pi_{J,k}(\theta,t^{1:S}|t_y)$ to maintain particle diversity.
In an alternative approach targeting $\pi_{M}(\theta|t_y)$,  \shortciteN{peters+fs08a} 
probabilistically  reject particles with weight below a given threshold.
The final form of the weight including the rejection mechanism involves the form (\ref{eqn:L3app}), and so 
is unbiased for all $S\geq 1$.

\subsubsection{Alternative population-based samplers}

\shortciteN{sisson+ft07}, \shortciteN{toni+wsis08} and \shortciteN{beaumont+cmr08} propose alternative population-based likelihood-free algorithms.
While deriving from different sampling frameworks, they are essentially the same sampler and utilise importance-sampling weights of the form (\ref{eqn:L3app}). Following the arguments in Section \ref{sec:IS}, such samplers successfully target both $\pi_M(\theta|t_y)$ and $\pi_J(\theta,t^{1:S}|t_y)$ for all $S\geq 1$, and produce identical algorithms.

\section{Discussion}
\label{section:discussion}

In this article, we have extended some existing likelihood-free samplers to incorporate multiple ($S>1$) 
auxiliary data generations, $t^{1:S}\in\mathcal{T}^S$. In doing so, we have established an ambiguity over the 
target distribution
of such samplers,
%
which
is problematic from an interpretative perspective.
Those algorithms targeting $\pi_M(\theta|t_y)$, and requiring estimates of likelihood ratios within acceptance probabilities or importance weights, require the number of Monte Carlo draws $S\rightarrow\infty$
 to avoid a theoretical bias.
 Fortunately, through an equivalence with a likelihood-free sampler targeting $\pi_{J}(\theta,t^{1:S}|t_y)$, inferences performed  with the marginal posterior sampler are in practice unbiased.
However, this practical unbiasedness does not justify the sampler targeting $\pi_M(\theta|t_y)$. Such samplers can only be theoretically justified from the perspective of the joint posterior $\pi_J(\theta,t^{1:S}|t_y)$ given by (\ref{eqn:jointS}) (c.f. \shortciteNP{delmoral+dj08}).
Alternative representations of likelihood-free models (e.g. \citeNP{wilkinson08}, \citeNP{fernhead+p10}) may not offer this interpretation.


It may be important for the future development of efficient likelihood-free samplers to make clear the distinction between the two different forms of target distribution. 
%
%
For instance, suppose that a future algorithm is sufficiently complicated that the usual joint posterior distribution strategy  (Section \ref{section:samplers}) of cancelling the intractable likelihood functions, $f(t|\theta)$, between the target and proposal distributions  is unavailable. The sampler must then be implemented with the marginal posterior target, $\pi_M(\theta|t_y)$,  via Monte Carlo integration. 
%
However, if this same algorithm also relies on the evaluation of ratios of likelihood estimates, then for finite $S$, this sampler may not be theoretically justified without further investigation.

\section*{Acknowledgments}

SAS and YF are supported by the ARC-DP scheme (DP0664970 and DP0877432). GWP is supported by APAS and CSIRO CMIS.
GWP thanks M. W\"{u}thrich for useful discussion, and ETH FIM and 
P. Embrechts for  financial assistance. 
MB would like to thank the UK MoD for funding through the DIF Defence Technology Centre.
This work  was partially supported by the NSF under Grant DMS-0635449
 to SAMSI.

\bibliographystyle{chicago}
\bibliography{abc-marginal-joint}

\begin{thebibliography}{}

\bibitem[\protect\citeauthoryear{Beaumont, Cornuet, Marin, and Robert}{Beaumont
  et~al.}{2009}]{beaumont+cmr08}
Beaumont, M.~A., J.-M. Cornuet, J.-M. Marin, and C.~P. Robert (2009).
\newblock Adaptive approximate {Bayesian} computation.
\newblock {\em Biometrika\/}~{\em 96}, 983--990.

\bibitem[\protect\citeauthoryear{Beaumont, Zhang, and Balding}{Beaumont
  et~al.}{2002}]{beaumont+zb02}
Beaumont, M.~A., W.~Zhang, and D.~J. Balding (2002).
\newblock Approximate {Bayesian} computation in population genetics.
\newblock {\em Genetics\/}~{\em 162}, 2025 -- 2035.

\bibitem[\protect\citeauthoryear{Blum}{Blum}{2010}]{blum09}
Blum, M. G.~B. (2010).
\newblock Approximate {Bayesian} computation: a non-parametric perspective.
\newblock {\em Journal of the American Statistical Association (to appear)\/}.

\bibitem[\protect\citeauthoryear{Bortot, Coles, and Sisson}{Bortot
  et~al.}{2007}]{bortot+cs07}
Bortot, P., S.~G. Coles, and S.~A. Sisson (2007).
\newblock Inference for stereological extremes.
\newblock {\em Journal of the American Statistical Association\/}~{\em 102},
  84--92.

\bibitem[\protect\citeauthoryear{{Del Moral}, Doucet, and Jasra}{{Del Moral}
  et~al.}{2006}]{delmoral+dj06}
{Del Moral}, P., A.~Doucet, and A.~Jasra (2006).
\newblock Sequential {Monte Carlo} samplers.
\newblock {\em J. R. Statist. Soc. B\/}~{\em 68}, 411 -- 436.

\bibitem[\protect\citeauthoryear{{Del Moral}, Doucet, and Jasra}{{Del Moral}
  et~al.}{2008}]{delmoral+dj08}
{Del Moral}, P., A.~Doucet, and A.~Jasra (2008).
\newblock An adaptive sequential {Monte Carlo} method for approximate
  {Bayesian} computation.
\newblock Technical report, University of Bordeaux.

\bibitem[\protect\citeauthoryear{Fernhead and Prangle}{Fernhead and
  Prangle}{2010}]{fernhead+p10}
Fernhead, P. and D.~Prangle (2010).
\newblock Semi-automatic approximate {Bayesian} computation.
\newblock Technical report, http://arxiv.org/abs/1004.1112.

\bibitem[\protect\citeauthoryear{Marjoram, Molitor, Plagnol, and
  Tavar\'e}{Marjoram et~al.}{2003}]{marjoram+mpt03}
Marjoram, P., J.~Molitor, V.~Plagnol, and S.~Tavar\'e (2003).
\newblock Markov chain {Monte Carlo} without likelihoods.
\newblock {\em Proc. Natl. Acad. Sci. USA\/}~{\em 100}, 15324 -- 15328.

\bibitem[\protect\citeauthoryear{Peters, Fan, and Sisson}{Peters
  et~al.}{2009}]{peters+fs08a}
Peters, G.~W., Y.~Fan, and S.~A. Sisson (2009).
\newblock On sequential {Monte Carlo}, partial rejection control and
  approximate {Bayesian} computation.
\newblock Technical report, UNSW. http://arxiv.org/abs/0808.3466.

\bibitem[\protect\citeauthoryear{Peters, Nevat, Sisson, Fan, and Yuan}{Peters
  et~al.}{2010}]{peters+nsfy10}
Peters, G.~W., I.~Nevat, S.~A. Sisson, Y.~Fan, and J.~Yuan (2010).
\newblock Bayesian symbol detection fr relay systems via likelihood-free
  inference.
\newblock {\em {IEEE} Transactions on Signal Processing\/}, in press.

\bibitem[\protect\citeauthoryear{Pritchard, Seielstad, Perez-Lezaun, and
  Feldman}{Pritchard et~al.}{1999}]{pritchard+spf99}
Pritchard, J.~K., M.~T. Seielstad, A.~Perez-Lezaun, and M.~W. Feldman (1999).
\newblock Population growth of human {Y} chromosomes: {A} study of {Y}
  chromosome microsatellites.
\newblock {\em Molecular Biology and Evolution\/}~{\em 16}, 1791--1798.

\bibitem[\protect\citeauthoryear{Ratmann, Andrieu, Hinkley, Wiuf, and
  Richardson}{Ratmann et~al.}{2009}]{ratman+ahwr08}
Ratmann, O., C.~Andrieu, T.~Hinkley, C.~Wiuf, and S.~Richardson (2009).
\newblock Model criticism based on likelihood-free inference, with an example
  in protein network evolution.
\newblock {\em Proc. Natl. Acad. Sci. USA\/}~{\em 106}, 10576--10581.

\bibitem[\protect\citeauthoryear{Reeves and Pettitt}{Reeves and
  Pettitt}{2005}]{reeves+p05}
Reeves, R.~W. and A.~N. Pettitt (2005).
\newblock A theoretical framework for approximate {Bayesian} computation.
\newblock In A.~R. Francis, K.~M. Matawie, A.~Oshlack, and G.~K. Smyth (Eds.),
  {\em Proceedings of the 20th International Workshop for Statistical
  Modelling, Sydney Australia, July 10-15, 2005}, pp.\  393--396.

\bibitem[\protect\citeauthoryear{Sisson and Fan}{Sisson and
  Fan}{2010}]{sisson+f10}
Sisson, S.~A. and Y.~Fan (2010).
\newblock Likelihood-free {Markov} chain {Monte Carlo}.
\newblock In S.~P. Brooks, A.~Gelman, G.~Jones, and X.-L. Meng (Eds.), {\em
  Handbook of {Markov} chain {Monte Carlo}}. Chapman and Hall/CRC.

\bibitem[\protect\citeauthoryear{Sisson, Fan, and Tanaka}{Sisson
  et~al.}{2007}]{sisson+ft07}
Sisson, S.~A., Y.~Fan, and M.~M. Tanaka (2007).
\newblock {Sequential Monte Carlo} without likelihoods.
\newblock {\em Proc. Natl. Acad. Sci.\/}~{\em 104}, 1760--1765. Errata (2009),
  106, 16889.

\bibitem[\protect\citeauthoryear{Tavar\'e, Balding, Griffiths, and
  Donnelly}{Tavar\'e et~al.}{1997}]{tavare+bgd97}
Tavar\'e, S., D.~J. Balding, R.~C. Griffiths, and P.~Donnelly (1997).
\newblock Inferring coalescence times from {DNA} sequence data.
\newblock {\em Genetics\/}~{\em 145}, 505 -- 518.

\bibitem[\protect\citeauthoryear{Toni, Welch, Strelkowa, Ipsen, and
  Stumpf}{Toni et~al.}{2009}]{toni+wsis08}
Toni, T., D.~Welch, N.~Strelkowa, A.~Ipsen, and M.~P.~H. Stumpf (2009).
\newblock Approximate {Bayesian} computation scheme for parameter inference and
  model selection in dynamical systems.
\newblock {\em J. R. Soc. Interface\/}~{\em 6}, 187--202.

\bibitem[\protect\citeauthoryear{Wegmann, Leuenberger, and Excoffier}{Wegmann
  et~al.}{2009}]{wegmann+le09}
Wegmann, D., C.~Leuenberger, and L.~Excoffier (2009).
\newblock Efficient approximate {Bayesian} computation coupled with {Markov}
  chain {Monte Carlo} without likelihood.
\newblock {\em Genetics\/}~{\em 182}, 1207--1218.

\bibitem[\protect\citeauthoryear{Wilkinson}{Wilkinson}{2008}]{wilkinson08}
Wilkinson, R.~D. (2008).
\newblock Approximate {Bayesian} computation ({ABC}) gives exact results under
  the assumption of model error.
\newblock Technical report, Univ. of Sheffield.

\end{thebibliography}

\end{document}